# Nano-Biotechnology: Structure and Dynamics of Nanoscale Biosystems

**Babu A Manjasetty [2]T.R.Gopalakrishnan Nair [3] Y. S. Ramaswamy**

[1] Research Professor, Structural Proteomics Platform, babu.manjasetty@gmail.com
[2] Director – RIIC, trgnair@ieee.org
[3] Research Professor, Nanotechnology Initiative, yagatiramu@gmail.com
[1,2,3] Research Industry Incubation Center, D S Institutions, Bangalore –78 India

**Abstract -** Nanoscale biosystems are widely used in numerous medical applications. The approaches for structure and function of the nanomachines that are available in the cell (natural nanomachines) are discussed. Molecular simulation studies have been extensively used to study the dynamics of many nanomachines including ribosome. Carbon Nanotubes (CNTs) serve as prototypes for biological channels such as Aquaporins (AQPs). Recently, extensive investigations have been performed on the transport of biological nanosystems through CNTs. The results are utilized as a guide in building a nanomachinary such as nanosyringe for a needle free drug delivery.

## I Introduction

Nanotechnology involves research and development on materials and species with dimensions of roughly 1 to 100 nm, where properties of matter differ fundamentally from those of individual atoms or molecules or bulk molecules. The term *nano* is derived from the Greek word meaning “dwarf”. In dimensional scaling, nano refers to $10^{-9}$, i.e., one billionth of a unit. Therefore, nanotechnology involves techniques and methods for studying, designing, fabricating and manipulating things at the nanoscale to understand and create materials, devices and systems to exploit these phenomena for novel applications [1].

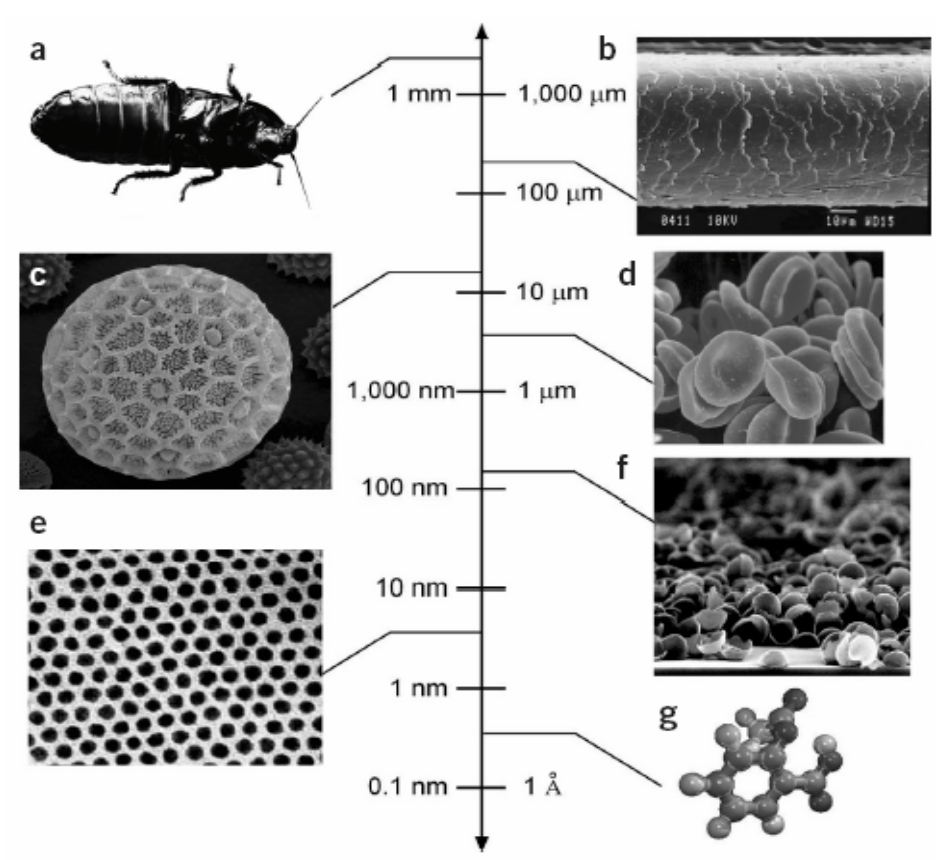

**Figure 1** Sizes of representative small objects. (a) A cockroach (b) A human hair (c) Pollen grain (d) Red blood cells (e) Cobalt nanocrystal superlattice (f) An aggregate of half-shells of palladium (g) Aspirin molecule

Two European scientists (Albert Fert, France and Peter Grunberg, Germany) were awarded the 2007 Nobel Prize in Physics for the discovery of a phenomenon called ‘*giant magnetoresistance*’. In this effect, very weak changes in magnetism generate larger changes in electrical resistance. This is how information stored magnetically on a hard disk can be converted to electrical signals that are read by a computer. This discovery revolutionized the computer industry by enabling computers, iPods and other digital devices store reams of data on ever-shrinking hard disks. Various morphologies of nanostructures such as belts, ribbons, cables, rods, tubes, rings, springs, helices, bows, tetrapods, spirals, needles and films have been synthesized and characterized. The research on nanostructures has rapidly expanded because of their unique and novel applications [2] in optics, optoelectronics, catalysis, biological sciences and piezoelectricity.

On the other hand, the determination of the human genome sequence is the most important achievement in the biotechnology sector [3, 4]. The full impact of this breakthrough in biomedical research will only be felt once the biological function of every gene product is known. This insight, in turn, requires a broad knowledge of the 3D structures of biosystems, since it is the structure that dictates biological function. At this juncture, the structural proteomics initiatives are being provided a comprehensive structural catalogue of nanoscale biosystems[5,6].

The interface of nanotechnology and biotechnology (nanobiotechnology) has produced tremendous applications in the domain of human health. Examples include the use of non-bleaching fluorescent nanocrystals (quantum dots) in place of dyes pulled glass capillaries with nanoscale tips utilized for injection of membrane-impermeable molecules (proteins, DNA). Another example is the use of carbon nanotubes (CNTs) for the delivery of biomolecular components to cells and smart biohybrid materials that enhance technologies by providing new avenues for regulating the activity of protein and DNA components.

This review article aims to provide an overview of recent developments in the context of nano biotechnology.

## II Structure, Dynamics and Function

### A. Structural Proteomics Approach

Three-dimensional structures are important for the functional analysis of proteins in a cell and for rational drug design [7]. The systematic study of protein structures, protein–protein, protein–nucleic acid and protein–small molecule interactions are crucial to understanding complex biological phenomena.

There have been important developments and milestones in the protein structure determination process. In the late sixties when protein crystallography was first established, the idea of determining the protein structures on a large scale was almost an impossible dream. Yet, 40 years later, large scale genome projects have provided the sequence infrastructure for protein analysis. Based on this availability of genome sequence data, the automated protein structure determination platforms have been established around the world by utilizing X-ray crystallography as a tool. This initiative was coined as, 'Structural proteomics'. The pilot study of these initiatives proved extremely successful at increasing the scope of structural science on protein families.

The ultimate goal of structural proteomics is to determine the structure of almost all the proteins in a cell or an organism. Structural proteomics approaches have led to thousands of protein structures being determined and deposited into the protein data bank (PDB)[9]. The representative novel protein structures determined at various structural proteomics centers are shown in Figure 2. The function of many of these novel structures remain unclear and can provide a large amount of tasks for structural proteomics as well as computational analysis for the next decade.

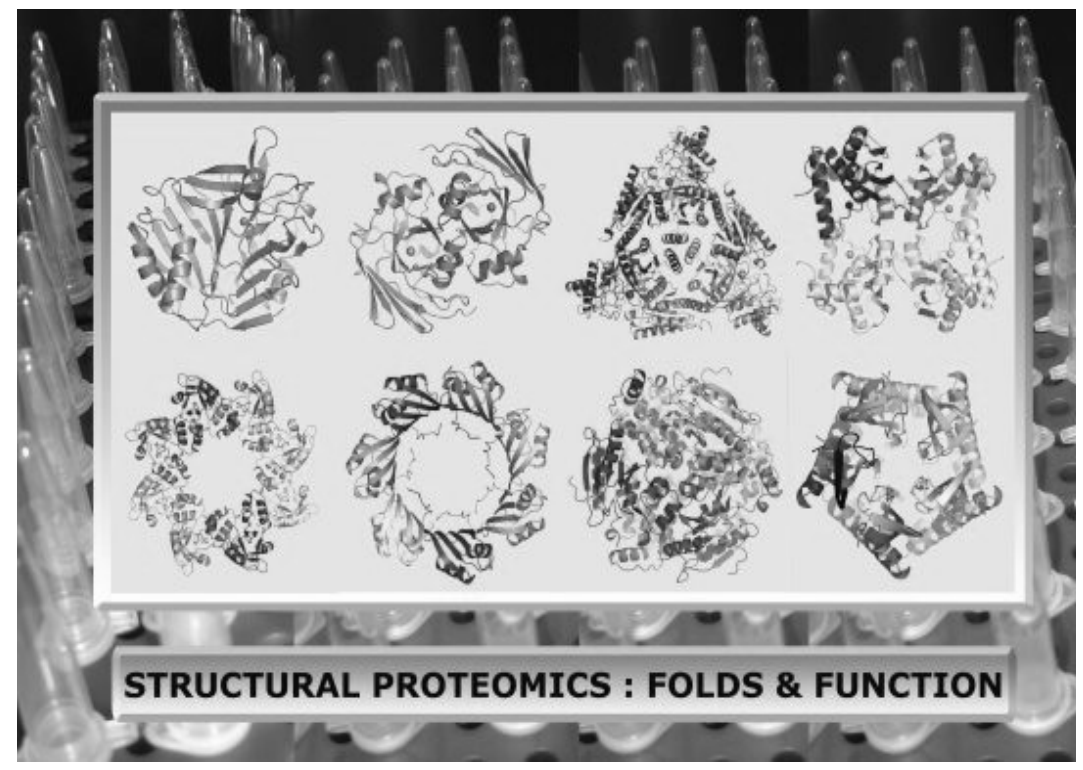

**Figure 2** The illustration depicts representative examples of novel protein structures determined at structural genomics centers, which adopt different oligomeric assemblies. To date, thousands of structures have been deposited into the Protein Data Bank (PDB) by structural genomics efforts. [8]

### B. Molecular Dynamics Approach

All-atom simulation is currently the most computationally demanding method for quantum dynamics and quantum calculations, in terms of computer power, communication speed, and memory load. Breakthroughs in electrostatic force calculation and dynamic load balancing have enabled molecular dynamics simulations of large biomolecular complexes [10].

Targeted molecular dynamics drive the system from one experimentally determined state to another, while simultaneously accounting for thermal fluctuations [11]. The simulations allow us to interpolate structures that occur between the two states during conformational changes important for the function of the biomolecular complex. These interpolated structures are then used to predict which regions of the complex are key to the biological function.

In the literature, the biomolecular dynamics simulations were originally simulated in 1980's on very short timescale dynamics (~ 10 ps) of small proteins (*bovine pancreatic trypsin inhibitor 'BPTI-1' ~500 atoms*) without solvent molecules due to limitations in computer power [12,13]. The advance in computer power allowed inclusion of solvent for these systems (*bovine pancreatic trypsin inhibitor 'BPTI-2' ~3100 atoms*). The availability of fast multiple algorithms further increased the simulation size to 1.26 $x10^4$ atoms (*photosynthetic reaction center of Rhodopseudomonas viridis, 'RHOD'*) [14], $2.4x10^4$ (*POPC lipid bilayer patch*) [15], $3.6x10^4$ atoms (*estrogen receptor binding domain plus DNA complex*) [16].

The fast multipole method was combined with a multiple-time-step method to simulate systems of $3.6x10^4$ atoms. A significant improvement in parallelization with dynamic load-balancing enabled simulation of solvated *acetylcholinesterase dimer ('AChE')* in the absence of long-range forces [17,18]. Finally, more sophisticated dynamic load balancing, replacing spatial domains by meta-domains based on compute-load, has produced simulations of $3.26x10^5$ atoms (*f1ATPase macromolecular complex*)[19].

The largest all-atom biomolecular simulation published to date is the *ribosome system* ($2.64x10^6$ atoms). It is approximately six times larger than the previous largest system simulated *dioleoyl phosphatidyl-choline (DOPC) bilayers* in which the formation of water filled pores can be followed at atomic resolution, and of a simpler system that allows detailed numerical analysis of the interactions of water with the electric field [20].

The ribosome is a giant molecular complex central to all living things and ancient in evolutionary terms. The job of the ribosome is to read genes and synthesize the proteins that the genes code for. The ribosome is a nano-scale computer that acts as the central information processing unit of the cell and is very much analogous to the central processing unit (CPU) of the cell. The ribosome is so fundamental to life that many parts of this molecular machine are identical in every organism ever sequenced, suggesting that these parts existed in their current form in the last common ancestor of life on earth. Until now, static snapshot structures of the ribosome have been available [21, 22].

Ribosome simulations have not only helped to elucidate a crucial molecular mechanism for gene expression, but have also opened the door for simulations of other large molecular machines important for gene expression and drug design [23]

## III Transport of Nanoscale Biosystems through Carbon Nanotube

Extensive studies have been carried out on carbon nanotubes (CNT) since its discovery and have revealed very interesting properties [24]. CNTs can be manufactured in various sizes, with diameters ranging from less than 1 nm to more than 100 nm. CNTs can attach to each other and form bundles by self-alignment [25]. CNTs found its applications in making nanoscale electronic devices or microscopic filters [26, 27].

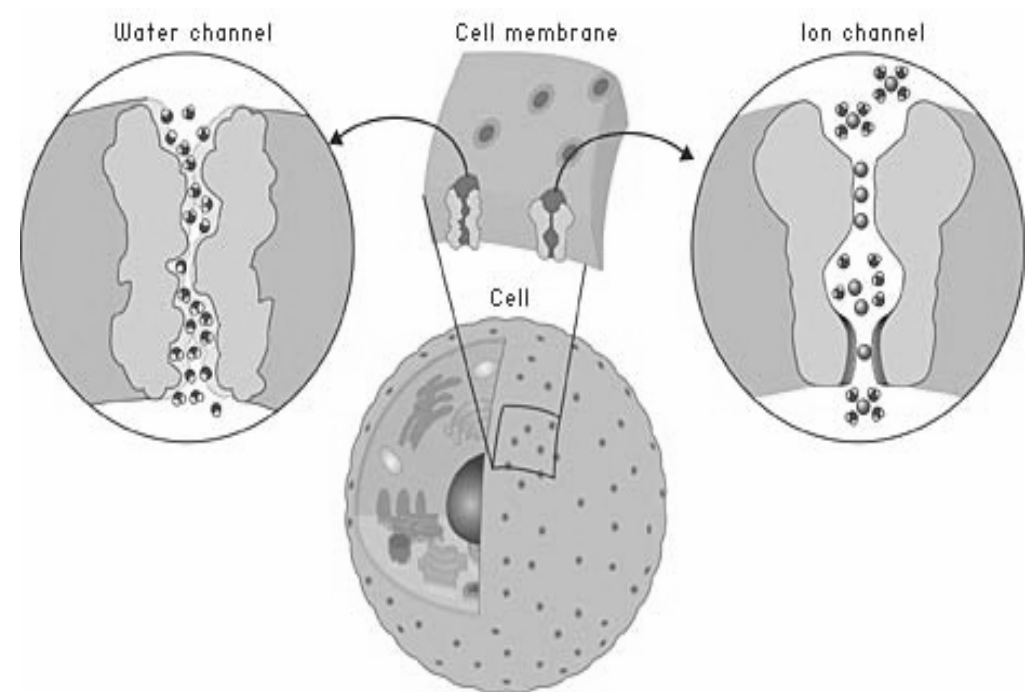

**Figure 3** The dividing wall between the cell and the outside world water channel (left) and an ion channel (right).

Computational studies have suggested that CNTs can be designed as molecular channels to transport water. A single-walled CNT, with a diameter of 81nm has been studied recently by molecular dynamics (MD) simulations [28]. The simulation studies revealed that the CNT was spontaneously filled with a single file of water molecules and that water diffused through the tube concertedly at a fast rate. This phenomenon is exactly analogous to the water channels in living cells as shown in figure 3 and 4.

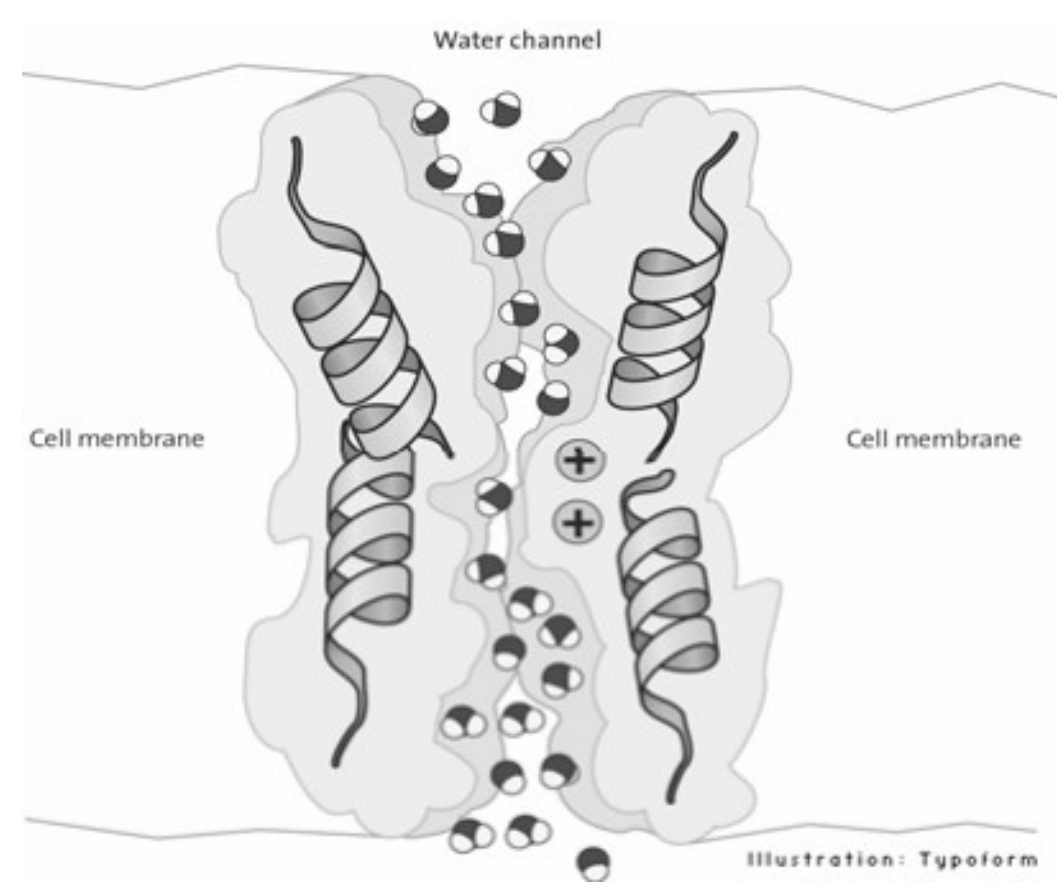

**Figure 4** Passage of water molecules through the aquaporin AQP. Due to the positive charge at the center of the channel, positively charged ions are deflected. This prevents proton leakage through the channel.

Most notable are aquaporins (AQPs), a family of membrane channel proteins, that are abundantly present in all kingdoms of life, including bacteria, plants, and mammals [29]. AQPs play critical roles in controlling the water contents of cells. More than ten different aquaporins have been found in the human body, and several diseases are connected to the impaired function of these channels. Crystal structure analysis suggests AQPs form tetramers in the cell membrane and facilitates transportation of water and, in some cases, other small solutes, across the membrane [30].

Further, molecular simulations studies have provided new insight into the mechanism underlying this fascinating property. Water molecules passing the channel are forced, by the protein's electrostatic forces, to flip at the center of the channel, thereby breaking the alternative donor-acceptor arrangement that is necessary for proton translocation [31].

Biological water channels are much more complex than CNTs, with irregular surfaces and highly inhomogeneous charge distributions. CNTs can serve as prototypes for these biological channels that can be investigated more easily by molecular dynamics simulations due to their simplicity, stability and small size. Many crucially important processes in biology involve the translocation of a nanobiosystem through nanometer-scale pores, such as DNA and RNA translocation across nuclear pores and protein transport through membrane channels. The study of interactions between CNTs and cellular components, such as membranes and biomolecules, is fundamental to the rational design of nanodevices interfacing with biological systems.

The water molecules flow through membranes of open-ended CNTs under an osmotic gradient and has been described in the earlier work [32]. In recent experiments, the extraordinarily fast transport of water in CNTs has been generally attributed to the smoothness of the CNT surface [33] [Figure 5]. Further, pressure-driven water flow through CNTs with diameters ranging from 1.66 to 4.99 nm is examined using molecular dynamics simulation [34].

An all-atom molecular dynamics simulation of RNA translocation through carbon nanotube membranes in explicit solvent is investigated. These simulations allow us to study translocation kinetics of RNA through the nanotube membranes at atomic detail [35].

Molecular dynamics simulation based on a novel polarizable nanotube model were performed to study the dynamics in translocation of a single-stranded deoxyribonucleic acid oligonucleotide through a polarized carbon nanotube membrane by an applied electric field [36]. Very recently, the CNT was used to transport a short RNA segment and it was found that the speed of translocation exhibits an exponential dependence on the applied potential differences. The RNA is transported while undergoing a repeated stacking and un-stacking process, affected by steric interactions with the membrane head groups and by hydrophobic interaction with the walls of the CNT[37].

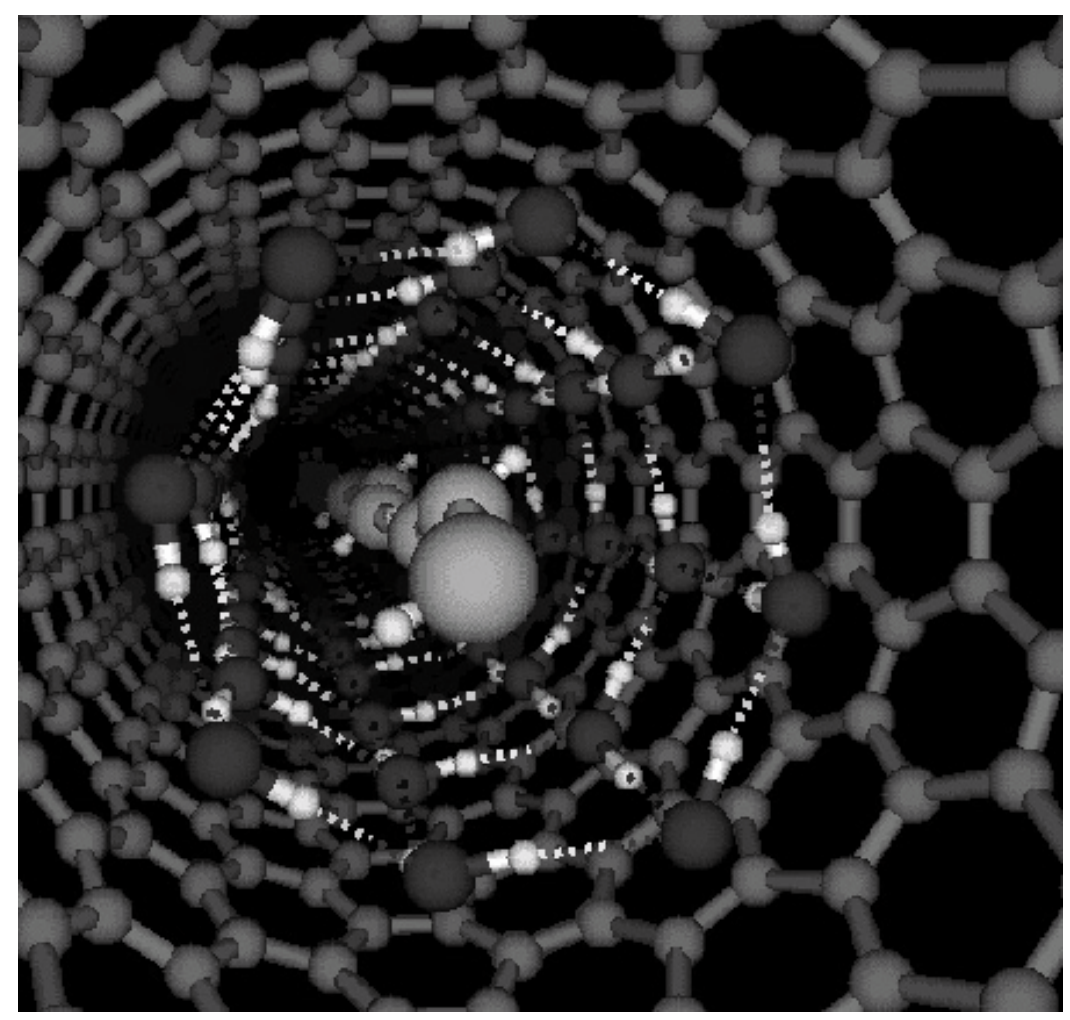

**Figure 5** The structure of nanotube-water model. The interior "chain" water molecules have been denoted by big spheres to distinguish them from the exterior "shell" water molecules denoted by small spheres.

The processes of molecular transport into cellular compartments through manufactured nanopores such as CNT is discussed in the context of applications in nanobiotechnology and nanomedicine. These results are a step closer to ultra-efficient, next-generation nanofluidic devices such as nanosyringe for drug delivery [38], water purification and nano-manufacturing applications.

## IV Conclusion

In a cell, molecules are often organized into functional aggregates, normally with nanometer dimensions. Visualizing and studying these structures-especially as they change dynamically during cycles of function,is one of the key challenges posed to nanobioscience. Information generated by X-ray crystallography coupled with molecular simulations has begun to clarify the dynamics of these nanostructures. Thishelps us design novel nanobiomachines for medical applications. Nanobiotechnology can bring a revolution in the design of delivery systems that are small and smart. Scientists are working towards the development of needle free drug delivery tools (nanoworms, nanopatches and nanosyringes).

## References

[1] Whiteside GM (2003) The 'right' size in nanobiotechnology. *Nature Biotechnology*, 21:1161-1165.

[2] Wang, ZL (2008) Oxide nanobelts and nanowires - growth, properties and applications. *J. Nanosci. Nanotechnol*. 8:27-55

[3] Lander ES, Linton LM, Birren B, Nusbaum C, Zody MC, et al., (2001) Initial sequencing and analysis of the human genome International Human Genome Sequencing Consortium. *Nature* 409:860-921.

[4] Venter JC, Adams MD, Myers EW, Li PW, Mural RJ, et al., (2001) The Sequence of the Human Genome. *Science*, 291:1304-1351.

[5] Heinemann U (2000) Structural genomics in Europe: slow start, strong finish. *Nat. Struct. Biol.,* 7:1940-1942.

[6] Manjasetty BA, Shi W, Zhan C, Fiser A and Chance MR (2007) A high through-put approach to protein structure analysis. Genetic engineering (New York), 28:105-127. Review.

[7] Zhang, C. & Kim, S. H. Overview of structural genomics: from structure to function. (2003) *Curr. Opin. Chem. Biol.* **7**, 28–32.

[8] Manjasetty BA, Turnbull AP, Panjikar S, Büssow K and Chance MR, "Automated technologies and novel techniques to accelerate protein crystallography for structural genomics". *Proteomics,* vol. 8, pp. 612 – 625, 2008.

[9] Berman HM, Westbrook J, Feng Z, Gilliland G, Bhat TN, Weissig H, Shindyalov IN and Bourne PE. (2000) The Protein Data Bank. *Nucleic Acids Research,* 28:235-242.

[10] Sanbonmatsu KY and Tung CS (2007) High performance computing in biology: multimillion atom simulations of nanoscale systems. *J. Struct. Biol*, 157:470-480.

[11] Schlitter, J., Engels, M. & Kruger, P. (1994). Targeted molecular dynamics: a new approach for searching pathways of conformational transitions. *J Mol Graph* 12, 84-9.

[12] Karplus, M. & McCammon, J. (2002) Molecular dynamics simulations of biomolecules. *Nature Structural Biology* 9, 646-652.

[13] McCammon, J. A., Gelin, B. R., Karplus, M. (1977). Dynamics of folded proteins. *Nature,* 267, 585-590.

[14] Heller, H., Grubmuller, H. & Schulten, K. (1990). *Molecular Simulation* **5**, 133-165.

[15] Board, J., Causey, J., Leathrum, J. & Schulten, K. (1992) Accelerated molecular dynamics simulation with the parallel fast multipole algorithm. *Chemical Physics Letters* 198, 89-94.

[16] Nelson, M., Humphrey, W., Gursoy, A., Dalke, A., Kale, L., Skeel, R. & Schulten, K. (1996) *International Journal of Supercomputer Applications and High Performance Computing,* 10, 251-268.

[17] Eichinger, M., Grubmuller, H., Heller, H. & Tavan, P. (1997). FAMUSAMM: An Algorithm for rapid evaluation of electrostatic interactions in molecular dynamics simulations. *Journal of Computational Chemistry* **18**, 1729-1749.

[18] Heffelfinger GS (2000) Parallel atomistic simulations. *Computer Physics Communications* 128, 219-237.

[19] Phillips, J., Gengbin, Z., Kumar, S. & Kale, L. (2002). NAMD: Biomolecular simulation on thousands of processors. *Proceedings of the SuperComputing 2002 annual meeting*.

[20] Tieleman, D. P. (2004) The molecular basis of electroporation. *BMC Biochem.,* 5:10.

[21] Ramakrishnan, V. (2002). Ribosome structure and the mechanism of translation. *Cell* **108**, 557-72.

[22] Tajkhorshid, E., Nollert, P., Jensen, M. O., Miercke, L. J., O'Connell, J., Stroud, R. M. & Schulten, K. (2002) Science, 296: 525-530.

[23] Sanbonmatsu KY, Joseph S and Tung CS (2005) Simulating movement of tRNA into the ribosome during decoding PNAS, 44:15854-15859.

[24] Iijima, S. (1991) Helical microtubules of graphitic carbon. Nature, 354:56–58.

[25] Dresselhaus, M. S., G. Dresselhaus, and P. C. Eklund. (1996) Science of Fullerenes and Carbon Nanotubes. Academic Press, San Diego, CA.

[26] Wind, S. J., J. Appenzeller, R. Martel, V. Derycke, and P. Avouris. (2002) Vertical scaling of carbon nanotube field-effect transistors using top gate electrodes. Appl. Phys. Lett. 80:3817–3819.

[27] Miller, S. A., V. Y. Young, and C. R. Martin. (2001) Electroosmotic flow in template-prepared carbon nanotube membranes. J. Am. Chem. Soc. 123:12335–12342.

[28] Hummer, G., J. C. Rasaiah, and J. P. Noworyta (2001) Water conduction through the hydrophobic channel of a carbon nanotube. Nature. 414:188–190.

[29] Borgnia, M., S. Nielsen, A. Engel, and P. Agre. 1999. Cellular and molecular biology of the aquaporin water channels. Annu. Rev. Biochem. 68:425–458.

[30] Morten Ø. Jensen, Tajkhorshid E and Schulten K. (2001) The mechanism of glycerol conduction in aquaglyceroporins. *Structure*, 9:1083-1093.

[31] Tajkhorshid E, Nollert P, Morten. Jensen, Larry J. W. Miercke, Joseph OÕ Connell, Stroud RM and Schulten K (2002) Control of the Selectivity of the Aquaporin Water Channel Family by Global Orientational Tuning, Science, 296:525-530.

[32] Kalra A, Garde S and Hummer G. (2003) Osmotic water transport through carbon nanotube membranes. PNAS, 100:10139-10140.

[33] Sony Joseph and N. R. Aluru (2008) Why Are Carbon Nanotubes Fast Transporters of Water? Nano Lett., 8, (2), 452–458.

[34] Thomas JA and McGaughey AJH (2008) Reassessing Fast Water Transport Through Carbon Nanotubes. Nano Lett., 8(9):2788-2793.

[35] In-Chul Yeh and Gerhard Hummer (2004) Nucleic acid transport through carbon nanotube membranes, PNAS, 101:12177-12182.

[36] Xie Y, Kong Y, Soh AK and Gao H (2007) Electric field-induced translocation of single-stranded DNA through a polarized carbon nanotube membrane. J.Chem.Phys., 127: 225101.

[37] Zimmerli U and Koumoutsakos P (2008) Simulations of Electrophoretic RNA Transport Through Transmembrane Carbon Nanotubes Biophysical Journal, 94: 2546-2557.

[38] Carlos F. Lopez,† Steve O. Nielsen, Preston B. Moore,‡ and Michael L. Klein (2004) Understanding nature's design for a nanosyringe, PNAS, 101:4431-4434.